\documentclass{kluwer}

\usepackage{graphicx}

\newcommand{\Hb}{\ensuremath{{\rm H}\beta}}
\newcommand{\Mgb}{\ensuremath{{\rm Mg}\, b}}

\newdisplay{guess}{Conjecture}

\begin{document}                                                                                   
\begin{article}
\begin{opening}        
\title{The Evolutionary Status of Early-type Galaxies \\
       in Abell 2390\thanks{Based on observations obtained at the Calar
       Alto Observatory, Spain.}}
\author{Alexander \surname{Fritz}, Bodo L. \surname{Ziegler}}
\institute{Universit\"atssternwarte G\"ottingen,
  Geismarlandstr. 11, D-37083, G\"ottingen, Germany,
  E-mail: afritz@uni-sw.gwdg.de}
\author{Richard G. \surname{Bower}} \author{Ian \surname{Smail}}
\author{Roger L. \surname{Davies}}
\institute{Department of Physics, University of Durham,
  Durham DH1\,3LE, UK}
\runningauthor{A. Fritz, B. L. Ziegler, R. G.\ Bower, I. Smail, \& R. L.\ Davies}
\runningtitle{The evolutionary status of early-type galaxies in Abell 2390}

\begin{abstract}
We explore the evolution of the early-type galaxy population in the rich
cluster Abell~2390 at $z=0.23$. For this purpose, we have obtained
spectroscopic data of 51 elliptical and lenticular galaxies with MOSCA at
the 3.5~m telescope on Calar Alto Observatory. As our investigation spans both
a broad range in luminosity ($-22.3\leq M_{B}\leq -19.3$) and a wide field of
view ($10'\times 10'$), the environmental dependence of different formation
scenarios can be analysed in detail as a function of radius from the cluster
center. In this paper, we present first results on the
Faber-Jackson relation and, for a subsample of 14 galaxies with morphological
and structural parameters from HST, we also investigate the evolution of the
Kormendy relation and the Fundamental Plane. We find a mild luminosity
evolution of the early-type galaxies in Abell~2390: our objects are on average
brighter by $\overline{m}_{B}\sim 0.4$~mag.
\end{abstract}

\end{opening}           

\section{Motivation and Sample Selection}

Early-type galaxies are tightly correlated via three parameters, the effective
radius $R_{{\rm e}}$ , effective surface brightness $\mu_{{\rm e}}$ and
velocity dispersion $\sigma$, that define a three dimensional parameter space
called the Fundamental Plane (FP) (\citeauthor{Dre:87} \citeyear{Dre:87},
\citeauthor{Djor:87} \citeyear{Djor:87}, \citeauthor{Ben:92} \citeyear{Ben:92}).
Projections of  this plane are the Faber-Jackson relation (FJR)
(\citeauthor{FJ76} \citeyear{FJ76}), luminosity $L$ {\it vs.} $\sigma$, and the
Kormendy relation (KR) (\citeauthor{Kor:77} \citeyear{Kor:77}),
$R_{{\rm e}}$ {\it vs.} $\mu_{{\rm e}}$. A study
of such scaling relations over different cosmic epochs offers powerful tests of
different aspects of galaxy evolution models and the hierarchical merging
scenario.

One of the key questions of early-type galaxy evolution is when and within what
timescales their stellar populations have been formed. Monolithic collapse
models predict a burst of star formation at high redshift
($z_{{\rm form}}\ge 2$) and a following passive evolution of the stellar
populations, whereas in the hierachical galaxy formation scenario the assembly
timescales for more massive galaxies are longer, resulting in somewhat younger
mean ages.

\begin{figure}[t]
\vspace*{0.3cm}
\tabcapfont
\centerline{%
\begin{tabular}{c@{\hspace{0.2cm}}c}
\includegraphics[width=0.48\textwidth]{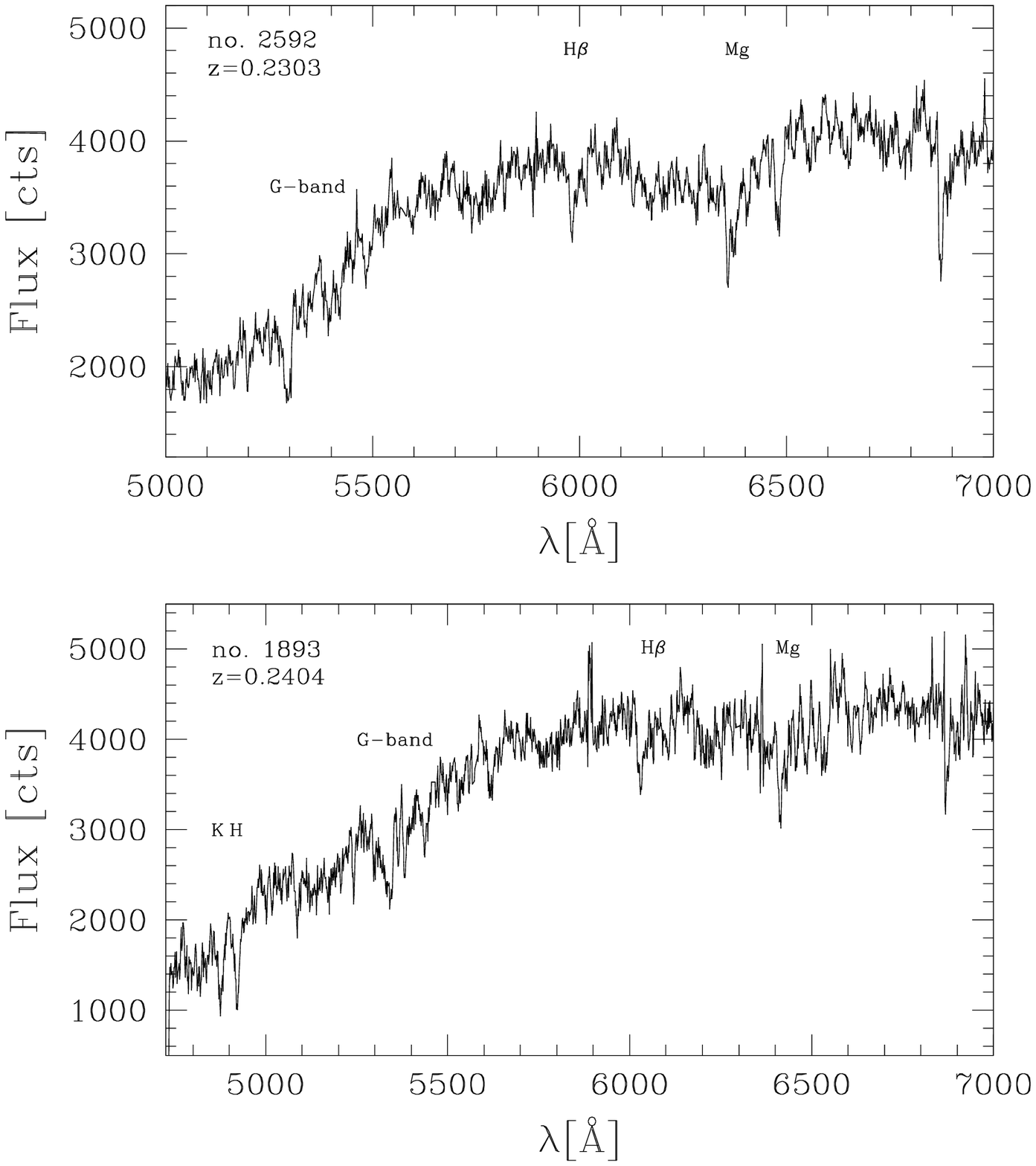} &
\includegraphics[width=0.48\textwidth]{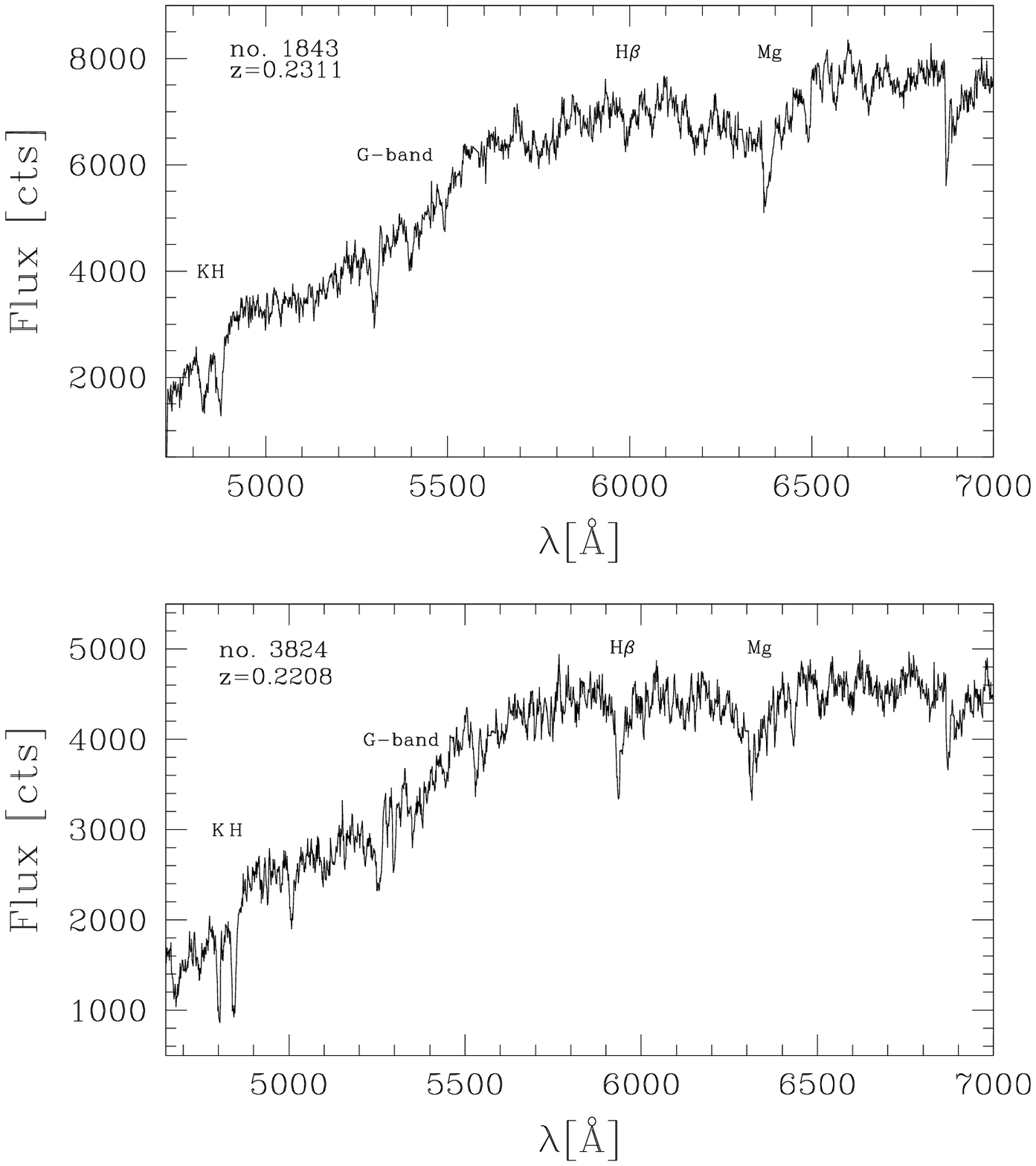} \\
\end{tabular}}
\caption{Typical observed A\,2390 early-type galaxy spectra. Not flux-calibrated
source counts are plotted against the observed wavelength. Prominent absorption
lines are marked.}
\label{spec22592}
\end{figure}

Previous spectroscopic studies were limited to a small number of the more
luminous galaxies.
To overcome bias and selection problems of small samples, we focus in this
study of the cluster Abell~2390 on a large number of objects ($N=51$), spanning
a wide range in luminosity $21.4<B<23.3$; $-22.4<M_{B}<-19.2$
(down to $M^{\ast}_{B}+1$) and a wide field of view of $\sim 10'\times 10'$
($2.5\times 2.5$~Mpc), analogous to the investigation of
Abell~2218 by \citeauthor{Zie:01} (\citeyear{Zie:01}). 
With this large sample, it is possible to explore variations in
early-type galaxy evolution as a function of distance from the cluster center
as well as for different sub-populations like S0 or Balmer-line enhanced
galaxies in a statistically significant way.

In this article a cosmological model with a deceleration parameter of
$q_{0}=0.1$ and a Hubble constant of $H_{0}$ = 65\,km\,s$^{-1}$\,Mpc$^{-1}$
is assumed.

\vspace*{-0.3cm}
\section{Observations}

During two observing runs (Sept. 1999 and July 2000) we gained a total of 51
different early-type galaxy spectra using the MOSCA spectrograph at the
3.5~m telescope at Calar Alto Observatory in Spain. The instrumental resolution
around \Hb\ and \Mgb\ ($5900\lsim\lambda\lsim 6400$~\AA) was 5.5~\AA\ FWHM,
corresponding to $\sigma_{\rm inst}\sim 100$~km~s$^{-1}$. 
The $S/N$ varies between 9.6 and 79.8 with an average value of $S/N\sim 41$.

The objects were selected on the basis of ground-based Gunn $i$-band images
(500~sec) obtained with the Palomar 5~m Hale telescope. Additional imaging
data from Mt. Palomar is available in the $U$ (3000~sec) and $B$ (500~sec)
filter bands and the WFPC2 camera onboard the
{\it Hubble Space Telescope (HST)} observed A\,2390 in the F555W and F814W
filter (10800~sec each), deep enough to determine structural
parameters down to $B_{{\rm rest}}\sim 23$ magnitudes
(\citeauthor{Z99} \citeyear{Z99}).

\section{Data Reduction and Analysis}

The reduction of the spectra was carried out using standard reduction
techniques implemented within MIDAS and IRAF. Particular care was taken of the
$S$-distortions of the MOSCA spectra. Examples
of final extracted 1-dimensional spectra are plotted in Fig. \ref{spec22592}.
Velocity dispersions were calculated with the Fourier Correlation Quotient
method described in \citeauthor{Ben:90} (\citeyear{Ben:90}).
Absolute magnitudes were calculated from our ground-based $UB\,i$
imaging. Using SExtractor object positions were determined and performing
aperture photometry apparent magnitudes were measured. After converting the
Gunn $i$ magnitudes to Cousins $I$, apparent magnitudes were transformed to
rest frame $B$ magnitudes with typical k-corrections of
${m}_{{\rm k-cor}}\approx 1.2$.
For those galaxies lying on {\it HST}\/ F814W images structural parameters were
determined by fitting the surface brightness profile with an $R^{1/4}$ and/or
an exponential law profile (\citeauthor{Sag:97} \citeyear{Sag:97}). A more
detailed description will be presented
in \citeauthor{Fri:02} (\citeyear{Fri:02}).

\begin{figure}
\vspace*{0.3cm}
\centerline{\includegraphics[width=11cm]{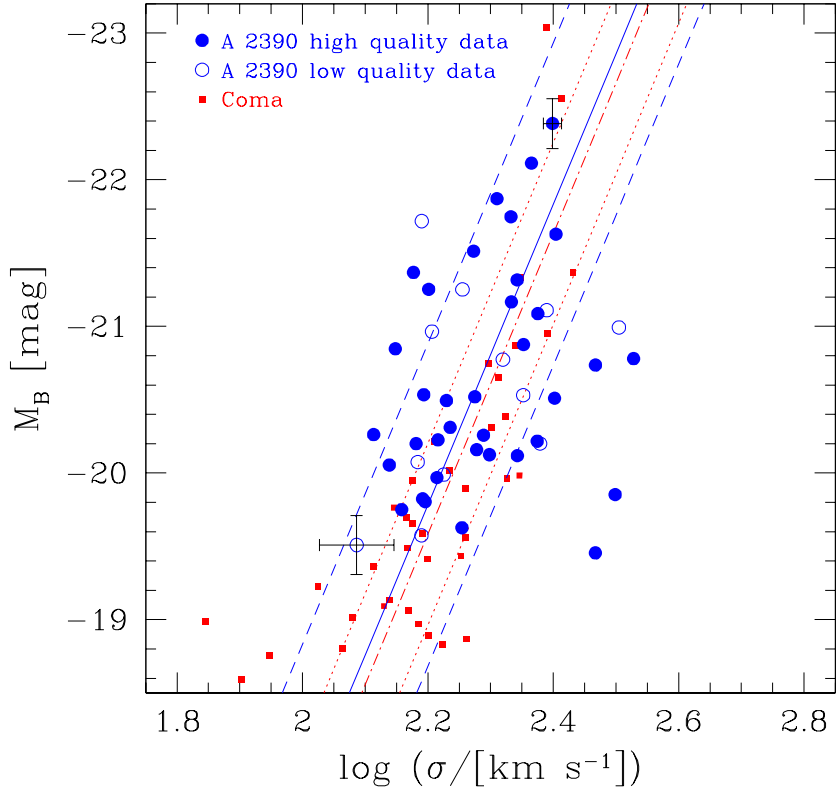}}
\caption[]{The $B$-band Faber-Jackson relation for A\,2390 compared to the Coma
sample of \citeauthor{Dre:87} (\citeyear{Dre:87}). The dot-dashed line shows
the bisector fit along with 1~$\sigma$ errors (dotted lines) for the local Coma
sample (filled squares). For the local bisector fit objects with
$M_{B}>-19.5$ and log~$\sigma <2.10$ were neglected. Under assumption of
constant slope a principal components fit (solid line) along 1~$\sigma$ errors
(dashed lines) for our A\,2390 sample is shown. The data are separated
in high (filled circles) and low (open circles) quality data
(see text for details). Typical error bars are plotted for two objects.
Restricting our sample to the high quality data the same amount of luminosity
evolution of $\overline{m}_{B}\sim 0.2$~mag as for the whole sample is found.}
\label{FJR}
\end{figure}

\section{Results}

The final $B$-band FJR is shown in Fig.~\ref{FJR}. In comparison to the local
Coma sample of \citeauthor{Dre:87} (\citeyear{Dre:87}) we find no
significant evolution, our magnitudes are brighter by 0.23~mag on average.
Depending on the accuracy of the velocity dispersion measurements, our galaxies
were divided into two subsamples, high quality and low quality data. The
quality parameter is a combination between various quantities, e.g.
the velocity dispersion ($\sigma$) measurement itself (including the error in
the single $\sigma$ and error in average $\sigma$),
$S/N$ of the spectrum and possible contamination by sky-lines. 

\begin{figure}[t]
\vspace*{0.3cm}
\tabcapfont
\centerline{\includegraphics[width=11cm]{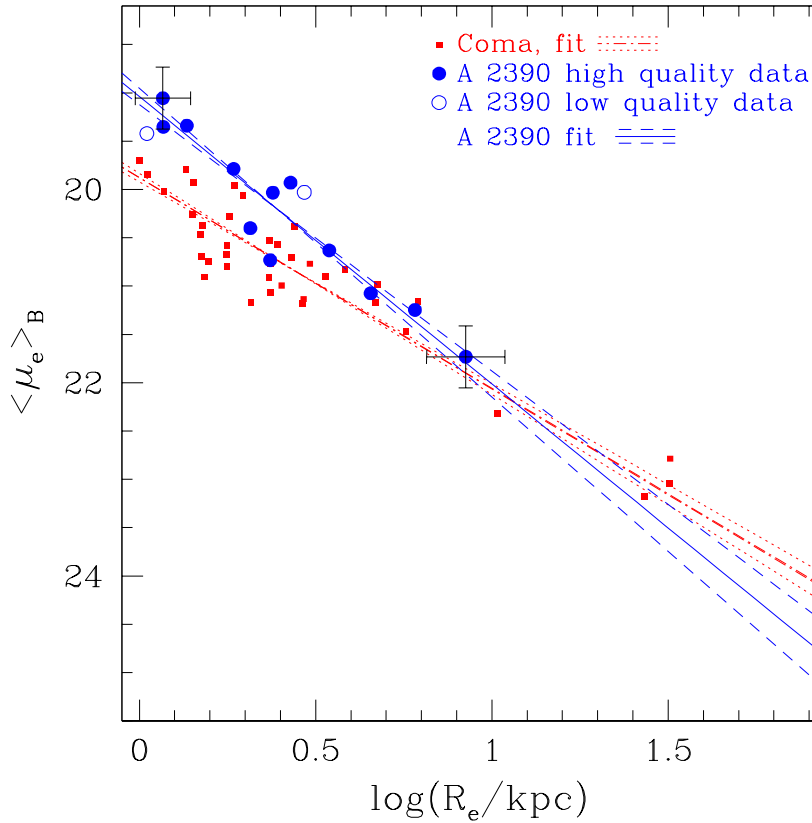}}
\caption{The $B$-band Kormendy relation for A\,2390 compared to the Coma
sample of Bender et al.~(1992). The dot-dashed line shows the local 100
iteration bootstrap bisector fit along with 1~$\sigma$ errors (dotted lines)
together with the bisector fit for our A\,2390 sample (solid line) with
1~$\sigma$ errors (dashed lines). Typical error bars are indicated
for two galaxies.}
\label{KR}
\end{figure}

In Fig.~\ref{KR} the KR for our {\it HST}\/ galaxies is shown. Comparing the
Coma sample of \citeauthor{Ben:92} (\citeyear{Ben:92}) with our data, we find a
similar result as with the FJR and again do not detect any significant
luminosity evolution. Our magnitudes are brighter by $\sim 0.5$ magnitudes
compared to the Coma objects.

\begin{figure}[t]
\vspace*{0.3cm}
\tabcapfont
\centerline{\includegraphics[width=11cm]{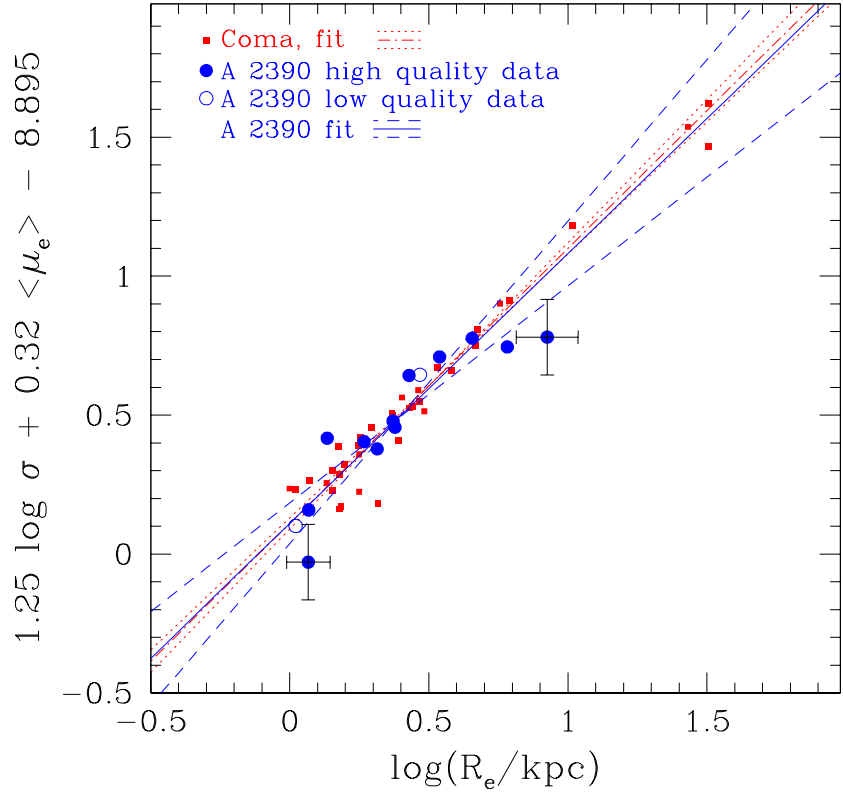}}
\caption{Edge-on view of the Fundamental Plane in restframe $B$ for A\,2390
compared to the Coma sample of Bender et al.~(1992). Symbol and line notations
are analogous to Fig.~\ref{KR}. An FP evolution of $0.48\pm 0.26$~mag
(1~$\sigma$) is found.}
\label{FP}
\end{figure}

The FP in restframe $B$ for our sample is plotted in Fig.~\ref{FP}.
Based on a smaller scatter than that of the KR and FJR, with the FP the
same result is seen (brighter by $\overline{m}_{B}\sim 0.5$~mag), which
indicates again only a modest luminosity evolution for A\,2390.
From these results we conclude that at a look-back time of $\sim 3.3$~Gyrs
most early-type galaxies of A\,2390 consist of an old stellar population and
that the formation redshift of elliptical galaxies must be at a much higher
redshift of about $z_{{\rm form}}\ge 2$, which is in agreement with the
formation picture of hierarchical merging in rich clusters
(\citeauthor{Kau:96} \citeyear{{Kau:96}}). \\ \\

\acknowledgements
It is a great pleasure to thank the organizers for an interesting and lively
meeting in the beautiful city of Porto.
AF was supported by a grant from the JENAM 2002 ``Galaxy Evolution" Workshop
organizers, through project ref. ESO/PRO/15130/1999 from FCT, Portugal.
We thank the Calar Alto staff for efficient observational support.
AF and BLZ ack\-nowledge financial support by the Volkswagen Foundation and
the Deutsche Forschungsgemeinschaft (DFG).

\end{article}
\end{document}